# Ferromagnetic state and phase transitions

Yuri Mnyukh
*76 Peggy Lane, Farmington, CT, USA, e-mail: yuri@mnyukh.com*
(Dated: June 20, 2011)

Evidence is summarized attesting that the standard exchange field theory of ferromagnetism by Heisenberg has not been successful. It is replaced by the crystal field and a simple assumption that spin orientation is inexorably associated with the orientation of its carrier. It follows at once that both ferromagnetic phase transitions and magnetization must involve a structural rearrangement. The mechanism of structural rearrangements in solids is nucleation and interface propagation. The new approach accounts coherently for ferromagnetic state and its manifestations.

## 1. Weiss' molecular and Heisenberg's electron exchange fields

Generally, ferromagnetics are spin-containing materials that are (or can be) magnetized and remain magnetized in the absence of magnetic field. This definition also includes ferrimagnetics, antiferromagnetics, and practically unlimited variety of magnetic structures. The classical Weiss / Heisenberg theory of ferromagnetism, taught in the universities and presented in many textbooks (*e. g.*, [1-4]), deals basically with the special case of a collinear (parallel and antiparallel) spin arrangement.

The logic behind the theory in question is as follows. There is a spontaneously magnetized crystal (*e. g.*, of Fe or Ni) due to a parallel alignment of the elementary magnetic dipoles. It remains stable up to its critical (Curie) temperature point when the thermal agitation suddenly destroys that alignment. It needed to be explained how the ferromagnetic state can be thermodynamically stable up to the really observed temperatures so high as 1042 K in Fe. It seemed unavoidable to suggest that the force holding the dipoles in parallel is the dipole interaction. Setting aside the probability that such interaction in Fe would rather cause mutual dipole repulsion than attraction, how strong must this interaction be? It followed from the Weiss' theory that it had to be about $10^4$ times stronger than the magnetic dipole interaction alone. The conclusion seemed undeniable: besides the magnetic dipole interaction, there is also interaction due to a much more powerful "molecular field" of unknown physical nature.

Heisenberg [5] accepted the Weiss' theory and developed its quantum-mechanical interpretation. His theory maintains that overlapping of the electron shells results in extremely strong *electron exchange interaction* responsible for collinear orientation of the magnetic moments. The main parameter in the quantum-mechanical formula was *exchange integral*. Its positive sign led to a collinear ferromagnetism, and negative to a collinear antiferromagnetism. Since then it has become accepted that Heisenberg gave a quantum-mechanical explanation for Weiss' "molecular field": "Only quantum mechanics has brought about explanation of the true nature of ferromagnetism" (Tamm [2]). "Heisenberg has shown that the Weiss' theory of molecular field can get a simple and straightforward explanation in terms of quantum mechanics" (Seitz [1]).

## 2. Inconsistence with the reality

General acceptance of the Heisenberg's theory of ferromagnetism remains unshakable to the present days. Judging from the textbooks on physics, one may conclude that it is rather successful [6]. In these books and other concise presentations every effort was made to portray it as basically valid and a great achievement, while contradictions, blank areas, and vast disagreements with experiment are either omitted as "details" or only vaguely mentioned. As a result, a new student gets wrong impression about the real status of the theory. In general, the theory remains basically unchallenged. But the more detailed the source is, the more drawbacks are exposed. There are experts who pointed out to its essential shortcomings.

Bleaney & Bleaney [7]: "There is no doubt that ferromagnetism is due to the exchange forces first discovered by Heisenberg, but the quantitative theory of ferromagnetism contains many difficulties".
"We have a broad understanding of the outlines of ferromagnetic theory, but not of the details. The exchange interaction between two electrons cannot be calculated *a priori* ... We cannot even be certain of its sign."

Belov [8]: "...Many important questions connected with the behavior of materials in the region [of ferromagnetic transition] remain unsettled or in dispute to the present time. These include ...the actual temperature behavior of the spontaneous magnetization near the Curie point, the causes of the 'smearing out' of the magnetic transition... the existence of 'residual' spontaneous magnetization above the Curie temperature, and the nature of the temperature dependence of elastic, electric, thermal, and other properties near the Curie point. It even remains unsettled what we should take to be the Curie temperature, and how to determine it".

"The theory of Weiss and Heisenberg cannot be applied to the quantitative description of phenomena in the neighborhood of the Curie point... Even for such a 'simple' ferromagnetic substance as nickel it is not possible to 'squeeze' the experimental results into the Weiss-Heisenberg theory".

Bozorth [3]: "The data for iron and for nickel [at low temperatures] show that the Weiss theory in either its original or modified form is quite inadequate".

"The Curie point is not always defined in accordance with the Weiss theory but in other more empirical ways..."

Crangle [9]: "It seems difficult to be convinced that direct exchange between localized electrons can be the main origin of the ferromagnetism in metals of the iron group".

Kittel [6]: "The Neel temperatures $T_N$ often vary considerably between samples, and in some cases there is large thermal hysteresis".

Feynman [10]: "Even the quantum theory deviates from the observed behavior at both high and low temperatures".

"The exact behavior near the Curie point has never been thoroughly figured out".

"The theory of the sudden transition at the Curie point still needs to be completed."

"We still have the question: why is a piece of lodestone in the ground magnetized?"

"To the theoretical physicists, ferromagnetism presents a number of very interesting, unsolved, and beautiful challenges. One challenge is to understand why it exists at all".

The last statement is especially indicative, considering that it was the primary purpose of the Weiss' and Heisenberg's theories to explain why ferromagnetism exists at all. Moreover, it turned out that the exchange forces, as powerful as they assumed to be, do not physically participate in the actual ferromagnetic phenomena. Thus, Seitz [1] maintained that the "Heisenberg's model…is too simple to be used for quantitative investigation of the real ferromagnetic materials". Tamm [2] noted that "it is the usual magnetic interaction of atoms [rather than exchange interaction] that is responsible for such, for example, phenomena as magnetic anisotropy and magnetostriction". In this respect many other phenomena could also be mentioned: domain structure, magnetic hysteresis, magnetocaloric effect, Barkhausen effect, first-order magnetic phase transitions, magnetization kinetics, and more. Remarkably, the question why the exchange forces do not exhibit themselves in those phenomena has never been raised.

There are also other phenomena and facts the exchange interaction offers no reasonable explanation, if at all. Among them:

(A) The value of the exchange integral for Ni was found lower by about two orders of magnitude needed to account for its Curie temperature.

(B) A collinear order of the atomic magnetic moments in ferro-, antiferro- and ferrimagnetics represents only particular cases, while there is, in fact, a great variety of non-collinear magnetic structures as well. The exchange field was unable to provide a parallel alignment in those innumerable magnetic structures.

(C) There are materials where magnetic moments are too far apart to make any direct exchange possible. The appropriate electron shells in the ferromagnetic rare-earth metals do not overlap. The 'exchange field' theory was expanded to those cases anyway, to become "superexchange".

(D) The actual speed of magnetization is well below of the theoretically expected.

(E) The exchange forces have the wrong sign.

## 3. The sign problem

Even the initial verifications of the Heisenberg's theory had to prevent its acceptance. The verifications have produced a *wrong sign* of the exchange forces. Feynman [10] was skeptical at least, as seen from these statements: "When it was clear that quantum mechanics could supply a tremendous spin-oriented force - even if, apparently, of the wrong sign - it was suggested that ferromagnetism might have its origin in this same force", and "The most recent calculations of the energy between the two electron spins in iron still give the *wrong* sign", and even "This physics of ours is a lot of fakery." The sign problem was later carefully examined in a special review [11] and found fundamentally unavoidable in the Heisenberg model. It was suggested that the "neglect of the sign may hide important physics."

## 4. Ferromagnetic phase transitions: from cooperative to magnetostructural

In order to present a coherent picture of ferromagnetism, which is the purpose of this article, the molecular mechanism of ferromagnetic phase transition should be established. With this in mind, it will be helpful to trace the evolvement of views on ferromagnetic phase transitions. Initially it was everyone's belief that they are of the second order - a cooperative phenomenon with a fixed (Curie) temperature of phase transition. Kittel [6] used Ni as an example to state: "This behavior classifies the usual ferromagnetic/paramagnetic transition as second order". In 1965 Belov wrote in his monograph "Magnetic Transitions" [8] that ferromagnetic and antiferromagnetic transitions are "concrete examples" of second-order phase transitions. His work was devoted to the investigation of spontaneous magnetization and other properties in the vicinity of the Curie points. The problem was, however, how to extract these "points" from the experimental data which were always "smeared out" and had "tails" on the temperature scale, even in single crystals.

Vonsovskii [4] was still on that initial stage when stated that the theory of second-order phase transitions provided an "impetus" to studies of magnetic phase transitions. But he already entered the second stage of the "evolvement" by recognizing that there are a number of the first-order ferromagnetic phase transitions. In his book about 25 such phase transitions were listed, still as rather "exotic". They were interpreted in the usual narrow-formal manner as those exhibiting abrupt changes and/or hysteresis of the magnetization and other properties. Some of these first-order ferromagnetic transitions Vonsovskii erroneously described as "apparent", where structural transitions occur before the ferromagnetic-to-paramagnetic transitions, but existence of genuine first-order ferromagnetic transitions was also recognized. The puzzling fact of their existence led to the numerous theoretical and experimental studies surveyed in the book. The conventional theory was in a predicament: the Curie point was not a point any more, and was rather a range of points and, even worse, was a subject to temperature hysteresis. Attempts were made, with no success, to complicate the theory by making the exchange field dependent on the lattice deformation, interatomic parameters, energy of magnetic anisotropy, *etc*. The first-order ferromagnetic phase transitions, so alien to the conventional theory, had to be accepted simply as an undeniable reality. *It was not realized that a first-order phase transition meant nucleation and growth, and not a critical phenomenon.*

The number of recognized first-order ferromagnetic phase transitions continued growing. They were found to be of the fist order even in the basic ferromagnetics - Fe, Ni and Co [12-14]. This process was accompanied by the increasing realization of structural changes involved. A new term *"magnetostructural"* transitions has come into use to distinguish them from not being "structural". At the present time the quantitative ratio "magnetostructural / second order" is dramatically shifting in favor of the "magnetostructural" phase transitions. The search with Google in June 8, 2011 produced
'second order ferromagnetic'….286,000 hits,
'first order ferromagnetic'...…..926,000 hits,
'magnetostructural transition'…718,000 hits.

## 5. The assumptions

The above trend is obvious, addressing us toward the conclusion that *all* ferromagnetic phase transitions are "structural", meaning they are always realized by nucleation and crystal rearrangements at the interfaces, rather than cooperatively. While this conclusion will formally remain our assumption, it is destined to be accepted as a fact. Designations of phase transitions as second order are always superficial. Not a single sufficiently documented example, ferromagnetic or otherwise, exists. This is because a nucleation-growth phase transition represents the most energy-efficient mechanism, considering that it needs energy to relocate only one molecule at a time, and not the myriads of molecules at a time as a cooperative process requires. Refer to [15].

The other assumption is: *the orientation of a spin is determined by the orientation of its atomic carrier*. Considering that the atomic carrier is an asymmetric entity, this simple assumption is more probable than ability of a spin to acquire different orientations in the same atom. These two assumptions represent the new fundamentals allowing to coherently account for ferromagnetic state and the numerous ferromagnetic phenomena. Knowledge of the actual molecular mechanism of nucleation-and-growth phase transitions will be necessary. Importantly, this will not require introduction of a "molecular field" of any kind in addition to the already existing chemical crystal bonding and magnetic dipole interaction..

## 6. The crucial part of crystal structure

Two opposing factors were considered by the Weiss' theory: the "molecular field" causing a parallel alignment of the ensemble of elementary magnets and the thermal agitation destroying this alignment. There the role of a crystal structure was implicitly reduced only to providing a positional, but not orientational, order to its magnetic dipoles. A system of atomic magnetic dipoles was a dipole system only. The objects of thermal agitation were the elementary magnets, and not the atoms carrying them. The *crystal field was overlooked*. There are powerful bonding forces combining molecules, ions, atoms, magnetic or not, into a crystal 3-D long-range order, both positional and orientational. *It is the crystal field that imposes one or another magnetic order by packing spin carriers in accordance with the structural requirements.*

## 7. The mechanism of nucleation-and-growth phase transitions

The following is a synopsis of the general mechanism of solid-state phase transitions and other structural rearrangements, deduced from the studies presented by the sequence of journal articles [16-29] and summarized in the book [13].

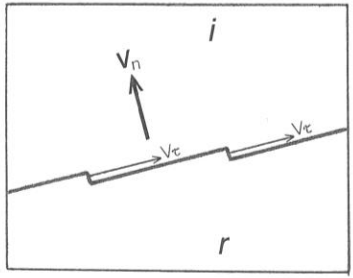

FIG. 1. The *edgewise* mechanism of phase transitions and any other rearrangements in solid state, such as at domain boundaries. The sketch illustrates the mode of advancement of interface in the **n** direction by shuttle-like strokes of small steps (kinks), filled by molecule-by-molecule, in the $\tau$ direction; *i* and *r* – are initial and resultant crystals, respectively. (A crystal growth from liquids is realized by the same manner). The kinks may consist of a single molecular layer or be a ladder-like conglomeration of smaller steps. Refer to [24,13] for more detailed description.

• Rearrangements in a solid state are a crystal growth by nucleation and propagation of interfaces. Neither ferromagnetic and ferroelectric phase transitions, nor phase transitions involving the orientation-disorder crystal (ODC) phase are excluded from this rule. Not a single sufficiently documented example exists of a transition being homogeneous (cooperative).
• The nuclei are located in specific crystal defects - microcavities of a certain optimum size. These defects contain information on the condition (*e.g.*, temperature) of their activation and orientation of the resultant crystal lattice. The nucleation can be epitaxial, in which case a certain orientation relationship between the initial and resultant structures is observed.
• The interface is a rational crystallographic plane of the resultant crystal lattice. It is named "contact interface" owing to a direct molecular contact between the two lattices without any intermediate layer. The molecular rearrangement proceeds according to *edgewise* (or *stepwise*) mechanism (Fig.1) involving formation of "kinks" (steps) at the flat interface and filling them, molecule-by-molecule, until the layer is complete, and building successive layers in this manner.

## 8. Accounting for ferromagnetism and its manifestations (including the problems cited by Feynman)

This will be done below within reasonable limits of a single article - mostly in a synopsis form.

♦ *Some problems are eliminated automatically*:
- There are two types of ferromagnetic phase transitions - second order and first order. (Only one exists).
- Application of the statistical mechanics to first-order ferromagnetic phase transitions. (Not applicable).
- The Curie point is blurred and subjected to hysteresis. (Phase transition temperature is not a Curie point).
- Magnetocrystalline (anisotropy) energy. (The partial impact of the crystal on spin directions is replaced by our premise that spin orientation is bound to the orientation of its carrier).

♦ *Stability of a ferromagnetic state*. (Feynman: "why ferromagnetism exists at all?"). Ferromagnetic state is a "slave" of crystal structure. A particular spin alignment ("magnetic structure") is determined by the requirements of crystal packing. The magnetic structure is an element of that 3-D packing, contributing a small positive or negative addition to the total crystal free energy. Ferromagnetism materializes in those cases when minimum free energy of the *crystal packing* requires placing spin carriers in the positions with their spins not mutually compensated. Despite of the possible destabilizing effect of the magnetic interaction, it is too weak to make any alternative crystal structure preferable. In brief: contribution of the magnetic interaction to the total crystal free energy is small as compared to that of crystal bonding; a ferromagnetic crystal is stable due to its low *total* free energy *in spite* of the destabilizing effect of the magnetic interaction.

♦ *"Why is a piece of lodestone in the ground magnetized?"* By razing this question, Feynman meant that, besides the stability problem, there must be an original cause turning non-ferromagnetic lodestone to ferromagnetic. Answer: it became ferromagnetic in the prehistoric times during its crystallization from liquid phase. The ferromagnetic state of lodestone is an inherent element of its crystal structure.

♦ *Existence of a great variety of non-collinear magnetic structures.* These are some types of magnetic structures in crystals: "simple ferromagnetic", "simple antiferromagnetic", "ferrimagnetic", "weakly ferromagnetic", "weakly non-collinear antiferromagnetic", "triangle", "simple helical", "ferromagnetic helical", and more. Only in the heavy metallic rare earths the following magnetic structures were listed [9]: "ferromagnet", "helix", "cone", "antiphase cone", "sinusoidally modulated", "square-wave modulated". The diversity in the mutual positions and orientations of spins can only be matched by the diversity in the world of crystal structures. This is not accidental: a magnetic structu re is *imposed* by the crystal, being secondary to the requirements of the crystal geometry.

♦ *Paramagnetic state.* It is usually assumed, as Weiss did, that the magnetic dipoles of the high-temperature phase of a ferromagnet lost their ferromagnetic alignment due to thermal rotation. The Weiss' view is understandable, for in his times the orientation-disordered crystals (ODC) were not yet discovered. The atoms and molecules, and not their spins alone, in the ODC state are engaged in a hindered thermal rotation. A zero magnetic moment of the high-temperature phase in question can also be not owing to the ODC state, but result from mutual compensation of its spins in the centrocymmetrical structure.

♦ *Ferromagnetic phase transitions.* A reorientation of spins involved in these phase transitions requires changing the orientation of spin carriers. The only way to achieve that is replacing the crystal structure. This occurs by nucleation and interface propagation. It follows that *all* ferromagnetic phase transitions are "magnetostructural". The term, however, is defective in the sense that it suggests existence of ferromagnetic phase transitions without structural change.

♦ *Magnetization by interface propagation.* The conventional theory does not explain why magnetization occurs in this manner rather than cooperatively in the bulk. Once again: magnetization is not a spin reorientation in the same crystal structure, but requires turning the spin carriers. The only way to turn the carriers is by crystal rearrangement. The mechanism of crystal rearrangements is nucleation and interface propagation. The possibility of a cooperative magnetization "by rotation" is thus ruled out. Refer to [31].

♦ *Magnetization "switching" and "reversal".* Their experimentally estimated ultimate speed in single-domain particles turned out three orders of magnitude lower than theoretically predicted [30]. The cause: whether they are activated by temperature, pressure, or external magnetic field, they always materialize by a relatively slow process of nucleation and propagation of interfaces. Refer to [31,32].

♦ *Origin of magnetic hysteresis.* The current theory was powerless to deal with magnetic hysteresis other than in a phenomenological manner, while its physical cause remained a question mark. *Solution***:** Magnetic hysteresis is a reflection of the *structural* hysteresis both in ferromagnetic phase transitions and in magnetization of domain systems. They require 3-D nucleation to begin and 2-D nucleation to proceed. The nucleation is heterogeneous, localized in specific defects – microcavities – where nucleation lags are encoded. These nucleation lags are the cause of magnetic hysteresis. Refer to [32].

♦ *Formation of magnetic hysteresis loops.* The "sigmoid" shape of the hysteresis loops is due to the balance between the increase in nucleation sites and the decrease in the amount of the original phase. Refer to [32].

♦ *Specific heat near the Curie transition.* (Feynman: "One of the challenges of theoretical physics today is to find an exact theoretical description of the character of the specific heat near the Curie transition - an intriguing problem which has not yet been solved. Naturally, this problem is very closely related to the shape of the magnetization curve in the same region"). *Solution*: The cooperative "Curie transition" does not exist. Solid-state phase transitions occur by nucleation and growth (Section 6). What believed to be a specific heat anomaly (called λ-anomaly) is not anomaly at all. It is the *latent heat* of a first-order phase transition (Fig. 2). Refer to [33] and Chapter 3 in [13].

♦ *Ferromagnetic domain structure*. An essential fact regarding ferromagnetic domain structure is that it is not specifically rooted in a ferromagnetic state, as Landau and Lifshitz [34] assumed. Domain structures are found also in antiferromagnetics, ferroelectrics, superconductors, organic crystals, etc. Their origin is *structural*. A ferromagnetic domain structure originates by multiple nucleation of the ferromagnetic phase in several equivalent structural orientations within the paramagnetic matrix. Growth of these nuclei and subsequent "magnetic aging" proceed toward minimizing the magnetic energy. Refer to [13], Sec. 2.8.6, 4.5 and 4.9.

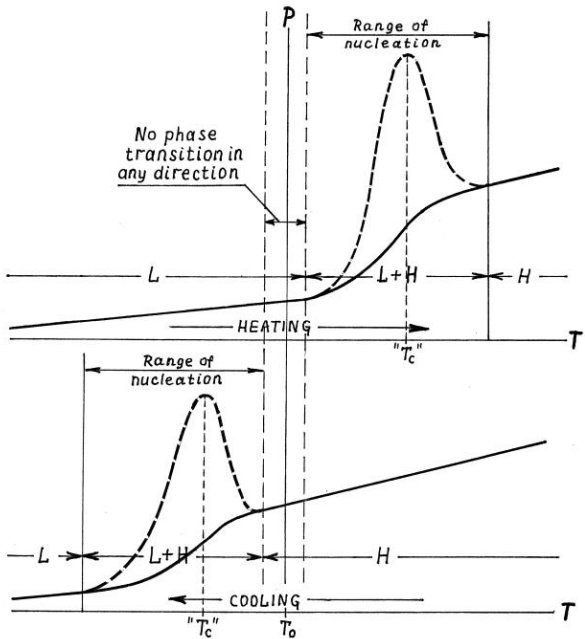

FIG. 2. The "anomalous" peaks of a physical property P, believe to be a heat capacity or magnetization, reside in the ranges of transition (actually, ranges of nucleation). The "critical (Curie) point $T_c$" at the λ-peak top (the common choice) is a subject of hysteresis, for there are two non-overlapping transition ranges, one above $T_o$ - for heating, and the other below $T_o$ - for cooling. In the adiabatic calorimetry these peaks are not a specific heat, but the *latent heat* of first-order (nucleation and growth) phase transitions. A differential scanning calorimetry would reveal the peak in a cooling run actually looking downward, being exothermic.

.♦ *Barkhausen effect* - short advances and stops during magnetization by magnetic field - is foreign to the traditional theory. The exchange field theory did not assume it. The domain theory may account only for the largest magnetization jumps, but they always consist of much smaller steps. The recent scientific work was devoted only to the phenomenological description of the effect, shedding no light on its nature [35]. But the effect is a direct manifestation of the crystal growth. In order to lower the crystal free energy in the applied magnetic field *H*, the spins of the ferromagnetic crystal have to turn toward the *H* direction, causing the structural rearrangement at the interfaces as shown in Fig. 1. Quick recrystallization of a whole layer at the domain boundary produces a magnetic "jump". The rearrangement of every successive layer is delayed by availability of next nucleus. The layers can be as thin as one lattice space, or they can be conglomerations of numerous elementary layers. In the latter case larger steps ("avalanches") appear on the magnetization curve.

A quick restructuring of a whole domain would produce the largest step, but it will inevitably consist of many smaller ones. Refer to [13], Sec. 4.10 and Addendum H.

♦ *Magnetostriction of Fe*. The phenomenon is not a kind of deformation, as usually believed. The α-Fe has a tetragonal rather than a cubic crystal structure. The magnetostriction results from the structural rearrangement, induced by application of magnetic field, that makes the direction of the longer crystallographic axis of the participated domains coincide with, or become closer to the direction of the applied magnetic field. Refer to [36].

♦ *Magnetocaloric effect.* It was acknowledged [37] that the "underlying physics behind the magnetocaloric effect is not yet completely understood". Now the physical nature of a "giant" magnetocaloric effect is explained in terms of the new fundamentals of phase transitions, ferromagnetism and ferroelectricity [13]. It is the *latent heat* of structural (nucleation-and-growth) phase transitions from a normal crystal state to the orientation-disordered crystal (ODC) state where the constituent particles are engaged in thermal rotation. The ferromagnetism of the material provides the capability to *trigger* the structural phase transition by application of magnetic field. Refer to [38].

♦ *Disparity with ferroelectricity.* Ferromagnetism and ferroelectricity are very similar phenomena with analogous set of manifestations. The standard theory was unable to find a unified approach to them since the Weiss/Heisenberg molecular field was applied only to ferromagnetism. No analog to it was found (or even needed) for ferroelectricity. Solution: This profound inconsistency disappears after the Weiss/Heisenberg molecular field is eliminated from consideration. Now the two phenomena have quite parallel explanations. Refer to [13].